\documentclass[twocolumn,aps,tightenlines,floatfix,showpacs]{revtex4}
\usepackage{amsmath}
\usepackage{graphicx}
\usepackage{hhline}
\usepackage{color}

\newcommand{\beq}{\begin{equation}}
\newcommand{\eeq}{\end{equation}}
\newcommand{\beqn}{\begin{eqnarray}}
\newcommand{\eeqn}{\end{eqnarray}}

\newcommand{\onbb}{$0\nu\beta\beta\,$}

\begin{document}

\title{Shell-model calculation of neutrinoless double-$\beta$ decay
of $^{76}$Ge}

\author{R.A.~Sen'kov$^{1}$ and M.~Horoi$^{2}$}

\affiliation{
$^{1}$ Department of Natural Sciences, LaGuardia Community College, CUNY, Long Island City, NY 11101, USA\\
$^{2}$ Department of Physics, Central Michigan University, Mount
Pleasant, Michigan 48859, USA
}


\begin{abstract}
In this article we present a more detailed version of our recent 
Rapid Communication  [Phys. Rev. C {\bf 90}, 051301(R) (2014)] where we calculate the nuclear matrix elements for neutrinoless double-$\beta$ decay of $^{76}$Ge. For the calculations we use a novel method that has perfect convergence properties and allows one to obtain the nonclosure nuclear matrix elements for $^{76}$Ge with a 1\% accuracy. We present a new way of calculation of the {\it optimal closure energy}, using this energy with the closure approximation provides the most accurate closure nuclear matrix elements. In addition, we present a new analysis of the heavy-neutrino-exchange nuclear matrix elements, and we compare occupation probabilities and Gamow-Teller strength with experimental data. 
\end{abstract}

\maketitle

\section{Introduction}

The search for neutrinoless double-$\beta$ decay is one of the most interesting and intensively studied topics of the modern nuclear physics. Neutrinos are unique particles, while there are many examples of truly neutral particles of integer spin (when the particle fully coincides with its antiparticle, for example, photon and $\pi^0$ meson), neutrinos are the only candidates for the truly neutral particles of half-integer spin. Explanation of such an asymmetry between the fermions and bosons is an ultimate challenge of the modern physics, and observation of neutrinoless double-$\beta$ decay would remove this difference and would make a significant contribution to our understanding of the Nature.  

Detecting neutrinoless double-$\beta$ (\onbb) decay is no doubts a very hard experimental task since the probabilities of \onbb decays are extremely small. Alongside with the experimental difficulties there are certain challenges in the theoretical part of the problem where accurate calculations of the nuclear matrix elements that involves the knowledge of a large number of nuclear states in the intermediate nucleus is required. Some of the recent theoretical attempts to address this problem within different approaches and models are: the quasiparticle random phase approximation (QRPA)~\cite{ves12,faess11,mika13}, the interacting shell model (ISM)~\cite{prl100,prc13}, the interacting boson model (IBM-2)~\cite{iba-2}, the generator coordinate method ~\cite{gcm}, and the projected hartree-fock bogoliubov model ~\cite{phfb}. 

The main target of all the approaches mentioned above is the calculation of the \onbb nuclear matrix elements (NMEs) that can be presented as a sum over the nuclear sates of the intermediate nucleus. In the case of $^{76}$Ge the intermediate nucleus is the odd-odd nucleus of $^{76}$As. One characteristic feature of most of the theoretical approaches is the use of the {\it closure}  approximation \cite{closure}, when the energies of the intermediate nuclear states are replaced with a constant value, so called closure energy $\langle E \rangle$. The great advantage of the closure approximation is that it allows one to analytically sum up over all the intermediate nuclear states by using the completeness relation. The disadvantage of this approximation is that the value of the closure energy is unknown and there is no any good way to calculate it. Moreover, one of the technical problems with the closure approximation is that the terms in the sum over the nuclear intermediate states have no unique sign, and there are positive and negative contributions of similar magnitudes in the sum. Thus varying the closure energy, even within a wide range of values, would not be able to adequately represent the true value of the nuclear matrix element. It should be noted though that at the current state of nuclear theory we cannot provide reliable calculations of many intermediate nuclear states, especially for odd-odd nuclei, so the closure approximation still plays a leading role in the \onbb nuclear matrix element calculations. 

In this paper, we summarize our recent progress in developing a shell-model based method of calculation of the \onbb NMEs beyond the closure approximation, the {\it mixed method} ~\citep{sh13,sh14}. We apply the mixed method to the calculation of the NMEs for \onbb decay of $^{76}$Ge, one of the most promising candidate for experimental observation of \onbb decay. The most sensitive limits on \onbb decay half-lives have been obtained from germanium-based experiments: the Heidelberg-Moscow experiment~\citep{hm01}, the International Germanium experiment~\cite{igex02}, and the GERDA-I experiment~\citep{gerda13}. $^{76}$Ge is the only isotope for which an observational claim has been made (though it was not accepted by the double-beta decay community)~\citep{hm04,hm06}. GERDA-II~\cite{gerda04} and MAJORANA DEMONSTRATOR \cite{mdem13}, the second generation of the germanium-based experiments, are in progress.

In the mixed method the low lying nuclear states of the intermediate nucleus are taken into account with their exact energies, both the wave functions and the energies are calculated using a shell model approach and a fine-tuned  effective shell model  Hamiltonian. For $^{76}$Ge it is impossible, and as we will show below, there is no need to calculate all the intermediate states because the intermediate states with the higher energies can be accounted in the closure approximation. Thus the mixed method has two free parameters: the cutoff parameter $N$ that separates the low lying states from the higher-energy states, and the closure energy that is only used for the contribution of the higher-energy states. 

The advantage of the mixed method is that the sensitivity of the mixed NMEs to the variation of the closure energy is significantly smaller than for the standard closure approximation  (see e.g. Fig. \ref{fig2} below). Also, the convergence properties of the NMEs as one increases the value of the cutoff parameter $N$ are incomparably better than if one considers only the low-lying intermediate states up to $N$ and does not include the higher-energy states (see Fig. \ref{fig1} below). Using the shell model, one of the most successful microscopic nuclear structure models, as the main tool of calculation brings in all the problems and challenges usually associated with the shell-model approach, namely  the restricted single-particle model space and the problem of getting a reliable effective shell model Hamiltonian. 

To calculate the NMEs of $^{76}$Ge we use NuShellX@MSU shell-model code~\cite{nushellxmsu}. The model space is $jj44$, which has as core $^{56}$Ni and the valence single-particle orbitals 
$f_{5/2}$, $p_{3/2}$, $p_{1/2}$, and $g_{9/2}$.
 We use JUN45 shell model Hamiltonian \cite{jun45}. Based on our experience with different nuclei, in order to achieve a reasonable accuracy for the NMEs calculations one needs to calculate a very small fraction of the intermediate states for each $J^\pi$ : {about 20 states or $^{48}$Ca ~\cite{sh13} and about 60 states for $^{82}$Se.} For the case of $^{76}$Ge we need only about 100 intermediate states in order to reach the necessary convergence. 

This paper presents an extensive analysis of the results recently published in short Rapid Communication \cite{sh2014}. It contains an extended analysis of the method used, it presents a number of new figures an tables that are used to clarify the results, and it contains refined versions of figures presented in Ref. \cite{sh2014}. In particular, we present $I$-pair decompositions for both light and heavy neutrino exchange NMEs that were recently used as a starting point to propose a new method of calculating these matrix elements \cite{brown-prl14}, and was recently used to make better estimates of the NMEs uncertainties \cite{BrownFangHoroi2015}. We also present the new way of calculation of the closure energies that can be used for the pure closure approaches, we argue that using our {\it optimal closure energies} with the standard closure approximation one can get the most accurate NMEs. We calculated the optimal closure energies for the \onbb decays of $^{48}$Ca, $^{82}$Se, and $^{76}$Ge isotopes. The effective Hamiltonian JUN45 was extensively validated and discussed in Ref. \cite{jun45}. Here we add to those observables studied in Ref. \cite{jun45} the neutron and proton occupancies in $^{76}$Ge and $^{76}$Se, and the Gamow-Teller strength in $^{76}$Ge.

\section{The nuclear matrix element}

Assuming the light-neutrino-exchange mechanism, the decay rate of a \onbb decay process can be written as \cite{ves12}
\beq \label{nme0}
\frac{1}{T_{1/2}} = G^{0\nu} | M^{0\nu} |^2 
\left(\frac{\langle m_{\beta \beta}\rangle}{m_e}\right )^2,
\eeq
where $ G^{0\nu} $ is the phase-space factor \cite{kipf12}, 
$ M^{0\nu} $ is the nuclear matrix element, $m_e$ is the electron mass, and $ \langle m_{\beta \beta}\rangle$ is the effective neutrino mass, which 
depends on the neutrino masses $m_k$ and the elements of neutrino mixing matrix $U_{ek}$ \cite{ves12},
\beq \label{eq2}
\langle m_{\beta \beta}\rangle = \left| \sum_k m_k U^2_{ek} \right|.
\eeq
The NME $M^{0\nu}$ is usually presented as a sum of three terms: Gamow-Teller ($M^{0\nu}_{GT}$), Fermi ($M^{0\nu}_{F}$), and Tensor ($M^{0\nu}_{T}$) NMEs (see, for example, Refs.~\cite{sh13}, \cite{sh14}, and \cite{prc10}), 
\beq \label{nme1}
M^{0\nu} = M^{0\nu}_{GT} - \left( \frac{g_{V}}{g_{A}} \right)^2  
M^{0\nu}_{F} + M^{0\nu}_{T}.
\eeq
Here we use $g_{A}=1.254$, for comparison with older results (using the modern $g_{A}=1.269$ would decrease the NME by less than 0.5\% \cite{sh14}), and $g_{V}=1$. 

\begin{figure*}
\centering
\begin{minipage}[t]{.43\textwidth}
\includegraphics[width=0.96\linewidth]{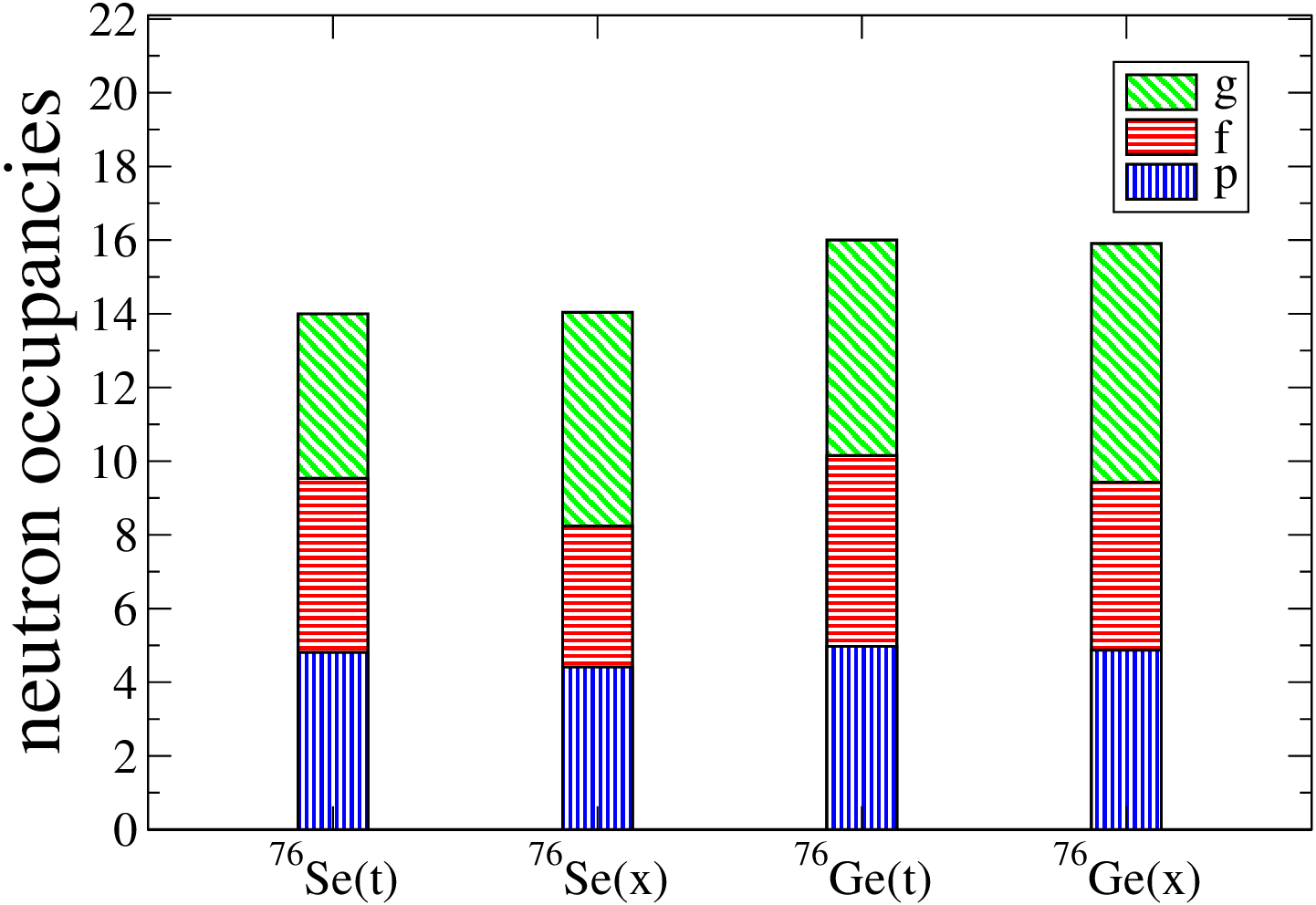}
\caption{(Color online) Theoretical (t) and experimental (x) neutron occupancies of the $p$ orbitals, $f_{5/2}$ orbital (f), and $g_{9/2}$ orbital (g) for $^{76}$Ge and $^{76}$Se. Data is taken from Ref. \cite{Schiffer2008}\\}
\label{fig-occn}
\end{minipage}\qquad
\begin{minipage}[t]{.43\textwidth}
\includegraphics[width=0.96\linewidth]{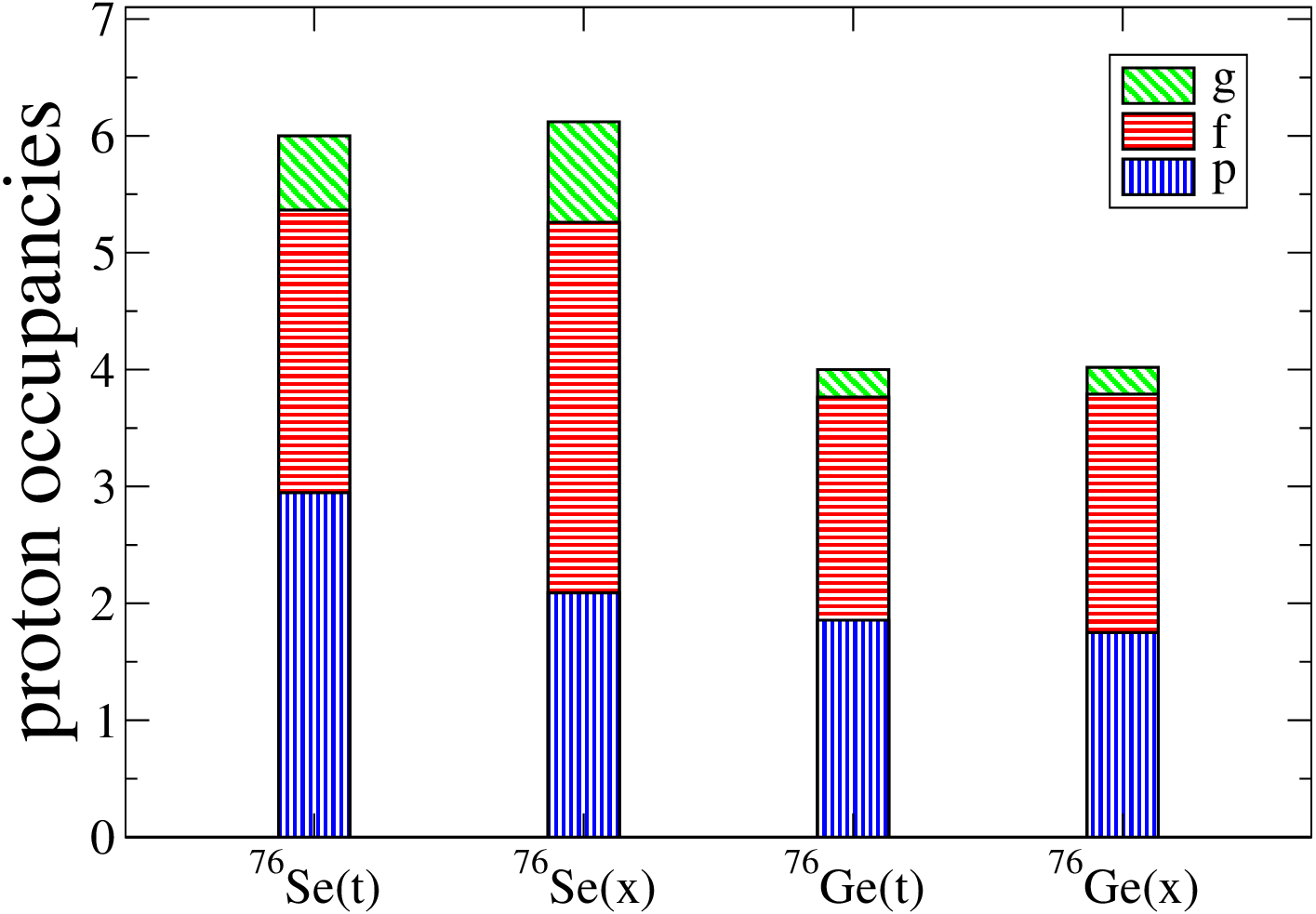}
\caption{(Color online) Same as Fig. \ref{fig-occn} for proton occupancies. Data is taken from Ref. \cite{Kay2009}\\}
\label{fig-occp}
\end{minipage}
\end{figure*}

In the case of \onbb decay of $^{76}$Ge, the matrix elements can be presented as an amplitude for the transitional process where the ground state $|i\rangle$ of the initial nucleus $^{76}$Ge changes 
into an intermediate state $|\kappa\rangle$ of the nucleus $^{76}$As 
and then to the ground state $|f\rangle$ of the final nucleus $^{76}$Se:     
\beq \label{nme2}
M^{0\nu}_{\alpha}=\sum_{\kappa} \sum_{1234} \langle 1 3 | 
{\cal O}_{\alpha} | 2 4\rangle
\langle f |  \hat{c}^\dagger_{3} {\hat{c}}_4 | \kappa \rangle
\langle \kappa |  \hat{c}^\dagger_{1} {\hat{c}}_2 | i \rangle.
\eeq
Here the sum over $\kappa$ spans all the intermediate states 
$|\kappa \rangle$, indices $1-4$ correspond to the single-particle 
quantum numbers, the label $\alpha$ describes different terms in the total NME (\ref{nme1}): Gamow-Teller ($\alpha=GT$), Fermi ($\alpha=F$), and Tensor ($\alpha=T$). The operators ${\cal O}_{\alpha}$ carry all the details of a \onbb decay process, they explicitly depend on the 
intermediate-state energy $E_\kappa$,
\beq \label{eq5}
{\cal O}_{\alpha}={\cal O}_{\alpha}(E_0+E_\kappa),
\eeq
through the energy denominators in perturbation theory. 
The actual form of the ${\cal O}_{\alpha}$ operators can be found in 
Ref.~\cite{sh13}. Here, we would like only to emphasize the energy 
dependence of these operators. The constant 
$E_0=\left[E_{gs}({}^{76}\mbox{As})-E_{gs}({}^{76}\mbox{Ge})\right] + 
Q_{\beta\beta}/2 \approx 1.943 \mbox{ MeV}$.

Exact calculation of the NMEs (\ref{nme2}) can be problematic due to 
the sum over a large number of intermediate states. One way to proceed in this situation is to restrict this sum by a state cutoff parameter $N$
\beq\label{nme4}
M^{0\nu}_{\alpha}(N)=\sum_{\kappa \leq N} 
\langle 1 3 | {\cal O}_{\alpha} | 2 4\rangle
\langle f |  \hat{c}^\dagger_{3} {\hat{c}}_4 | \kappa \rangle
\langle \kappa |  \hat{c}^\dagger_{1} {\hat{c}}_2 | i \rangle,
\eeq
here and below the sum over the repeated indexes 1,2,3, and 4 is assumed. In this {\it running nonclosure} approach, the NMEs defined by Eq. (\ref{nme4}) depend on the cutoff parameter $N$, they reach the exact values (\ref{nme2}) when $N \rightarrow \infty$: $M^{0\nu}_{\alpha} \equiv M^{0\nu}_{\alpha}(\infty)$. Success of the running nonlcosure approach is defined by the convergence properties of $M^{0\nu}_{\alpha}(N)$ as a function of $N$. 

Another way to proceed in this situation is to use the {\it closure} approximation. In the closure approximation the energies of intermediate states are replaced by a constant value as
\beq \label{eq7}
\left\{ 
\begin{array}{l}
E_0+E_\kappa \rightarrow \langle E \rangle,  \\
{\cal O}_\alpha (E_0+E_\kappa) \rightarrow \tilde{{\cal O}}_\alpha \equiv {\cal O}_\alpha(\langle E \rangle),
\end{array}
\right.
\eeq 
where $\langle E \rangle$ is the closure energy. Values of $\langle E \rangle$ from Ref. \cite{tomoda} are frequently used.

We introduce two forms of the closure approximation: the {closure} (or pure closure) and the {running closure} approximations \cite{sh14}. The {\it running closure} NMEs is presented similarly to the running nonclosure nuclear matrix elements (\ref{nme4}):
\beq\label{nme5}
{\cal M}^{0\nu}_{\alpha}(N)=\sum_{\kappa \leq N} 
\langle 1 3 | \tilde{{\cal O}}_{\alpha} | 2 4\rangle
\langle f |  \hat{c}^\dagger_{3} {\hat{c}}_4 | \kappa \rangle
\langle \kappa |  \hat{c}^\dagger_{1} {\hat{c}}_2 | i \rangle.
\eeq
${\cal M}^{0\nu}_{\alpha}(N)$ depend on both the state cutoff parameter $N$ and on the closure energy $\langle E \rangle$, when $N \rightarrow \infty$ the running closure NMEs reach their {\it closure} values  
\beq \label{nme3}
{\cal M}^{0\nu}_{\alpha}\equiv {\cal M}^{0\nu}_{\alpha}(\infty)=
\langle 1 3 | \tilde{{\cal O}}_{\alpha} | 2 4\rangle
\langle f |  \hat{c}^\dagger_{3} {{c}}_4 \hat{c}^\dagger_{1} {\hat{c}}_2 
| i \rangle,
\eeq
where we could remove the sum over intermediate states in Eq. (\ref{nme5}) using the completeness relation 
$ \sum |\kappa\rangle \langle \kappa | = {\hat I}$. Equation (\ref{nme3}) presents the standard closure approximation -- the simplest and commonly used method for \onbb decay NMEs calculations. 
The closure NMEs (\ref{nme3}) depend on the closure energy $\langle E\rangle$ which is not known and can not be calculated, which brings an uncertainty of about 10\% in the NMEs (see, for example, \cite{sh13,sh14,prc10}). 

\begin{figure}
\includegraphics[width=0.72\linewidth,angle=-90]{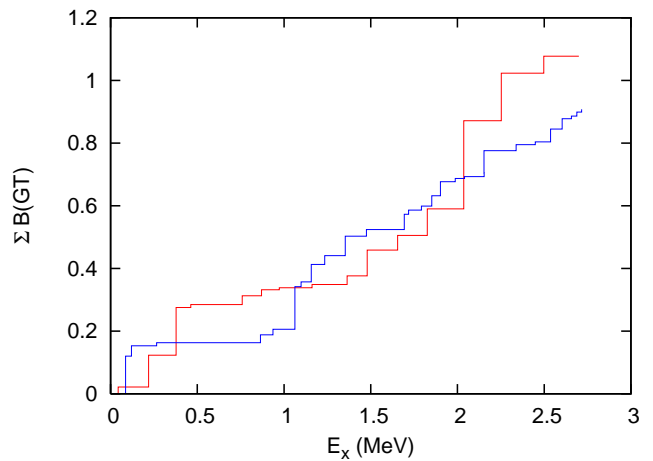}
\caption{(Color online) The running sum of the Gamow-Teller strength in $^{76}$Ge: red line shows the calculated sum and the blue line is based on the high-resolution charge-exchange data \cite{Thies2012}. }
\label{fig-gtd}
\end{figure}

In some cases, for example, the \onbb decay of $^{48}$Ca, the running nonclosure NMEs converge pretty fast and matrix elements can be computed within the standard shell model approach \cite{sh13}. However the running nonclosure approach cannot be directly used for the heavier cases, such as \onbb decay of $^{82}$Se and $^{76}$Ge, where only a few hundred intermediate states can be calculated. 

To resolve this problem the {\it mixed} (or just {\it nonclosure}) method was introduced \cite{sh13,sh14}. The mixed NMEs are presented as the following combination of the running nonclosure, closure, and running closure NMEs
\beq\label{nme6}
{\bar M^{0\nu}_{\alpha}}(N)=
{M}^{0\nu}_{\alpha}(N)
+{\cal M}^{0\nu}_{\alpha}
-{\cal M}^{0\nu}_{\alpha}(N).
\eeq 
In the mixed method the intermediate states below the cutoff parameter $N$ are taken into account by the first nonclosure term ${M}^{0\nu}_{\alpha}(N)$ and the states above the $N$ are included within the closure approach by $\left[{\cal M}^{0\nu}_{\alpha} - {\cal M}^{0\nu}_{\alpha}(N) \right]$. It was shown that the mixed NMEs (\ref{nme6}) converge significantly faster than the running matrix elements separately. It was also shown that the mixed NMEs have much weaker dependence on the closure energy $\langle E \rangle$ compared with the closure NMEs \cite{sh13,sh14}.

\begin{figure}
\begin{center}
\includegraphics[width=0.47\textwidth]{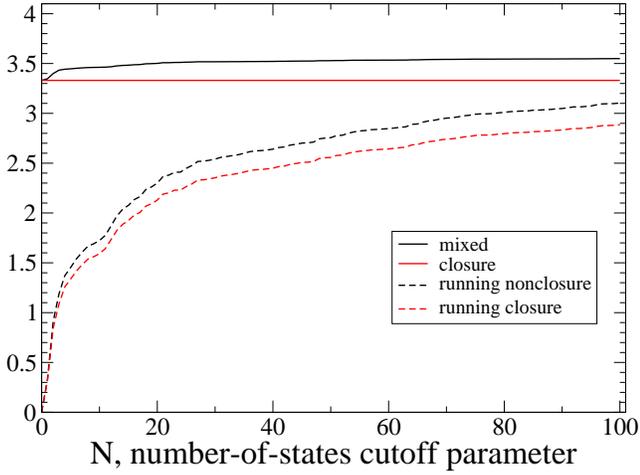}
\caption{(Color online) Convergence of NMEs (light-neutrino exchange) 
as a function of the cutoff parameter $N$ 
calculated with different approximations: mixed (black solid curve),  closure (red solid curve), running nonclosure (black dashed curve), and running closure (blue dashed curve). All calculations were done with CD-Bonn SRC and $\langle E \rangle=9.41$ MeV \cite{tomoda}.\\
}\label{fig1}
\end{center}
\end{figure}

The nonclosure approach allows one to calculate the \onbb decay NMEs for a fixed spin and parity $J^\pi$ of the intermediate states $|\kappa\rangle$,
\beq \label{jdec}
M^{0\nu}_{\alpha}(J)=\sum_{\kappa, \; J_\kappa = J} 
\langle 1 3 | {\cal O}_{\alpha} | 2 4\rangle
\langle f |  \hat{c}^\dagger_{3} {\hat{c}}_4 | \kappa \rangle
\langle \kappa |  \hat{c}^\dagger_{1} {\hat{c}}_2 | i \rangle,
\eeq 
where the sum over $\kappa$ spans all the intermediate states with a given spin and parity $J^\pi$. This $J$ decomposition can be obtained only within a nonclosure approach. Another way to decompose NMEs of a \onbb decay process is associated with the closure approximation. In this decoupling scheme the single-particle states $|1\rangle$ and $|3\rangle$ (proton states) and the states $|2\rangle$ and $|4\rangle$ (neutron states) in the two-body matrix elements $\langle 1 3 | {\cal O}_\alpha | 2 4 \rangle$ are coupled to certain common spin $I$
\beq \label{idec}
M^{0\nu}_{\alpha}(I)=\sum_{\kappa}
\langle 1 3, I | {\cal O}_{\alpha} | I, 2 4 \rangle
\langle f |  \hat{c}^\dagger_{3} {\hat{c}}_4 | \kappa \rangle
\langle \kappa |  \hat{c}^\dagger_{1} {\hat{c}}_2 | i \rangle,
\eeq 
here the sum over intermediate states is not restricted (for the details see Ref. \cite{sh13}). The total matrix elements can be obtained using any of these decoupling schemes as   
\beq \label{nme7}
M^{0\nu}_\alpha=\sum_{J} M^{0\nu}_\alpha(J)=\sum_{I} M^{0\nu}_\alpha(I).
\eeq 

We also analyze the NMEs for the right-handed 
heavy-neutrino-exchange mechanism, whose corresponding contribution to the total decay rate can be written as
\beq\label{hnme0}
\left[ T^{0\nu}_{1/2} \right]_{\mbox{heavy}}^{-1} = G^{0\nu} | M^{0\nu}_N |^2
|\eta_{NR}|^2,
\eeq
where the heavy-neutrino-exchange matrix elements $M^{0\nu}_N$ have a 
structure similar to that of the light-neutrino-exchange NMEs, while the parameter $\eta_{NR}$ depends on the heavy-neutrino masses (for more details see, for example, Ref. \cite{prc13}). One difference between the heavy- and the light-neutrino-exchange mechanisms is that the heavy-neutrino-exchange NMEs do not depend on the energy of intermediate states. Thus for the heavy-neutrino-exchange mechanism the closure approach provides the exact matrix elements. 

\begin{figure}
\begin{center}
\includegraphics[width=0.46\textwidth]{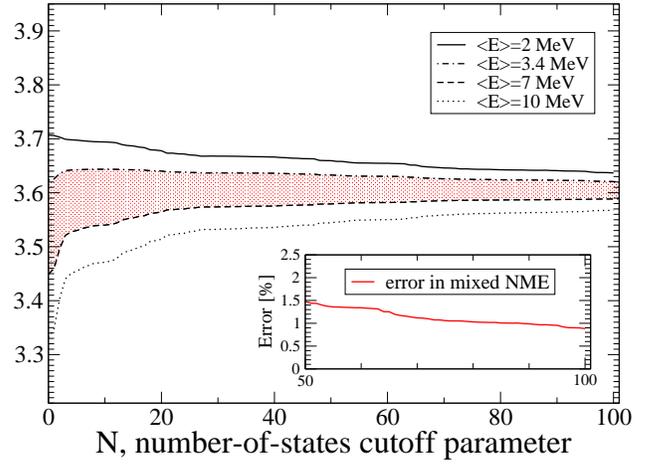}
\caption{ (Color online) Dependence of mixed NMEs (light-neutrino exchange) on the cutoff parameter $N$ calculated for different average closure energies $\langle E \rangle$. The main panel: $\langle E \rangle=2$ MeV (solid curve), $\langle E \rangle=3.4$ MeV  (dash-dotted curve), $\langle E \rangle=7$ MeV (dashed curve), and $\langle E \rangle=10$ MeV (dotted curve). 
The insert shows the uncertainty in the value of mixed NMEs corresponding to the shaded area from the main panel.\\}\label{fig2}
\end{center}
\end{figure}

\section{Nuclear structure calculations}\label{structure}

As we mentioned in the introduction, we use a shell model approach to calculate the NMEs for $^{76}$Ge. The valence space used here is $jj44$, which has as core $^{56}$Ni and the active single-particle orbits $f_{5/2}$, $p_{3/2}$, $p_{1/2}$, and $g_{9/2}$. A reliable effective shell model Hamiltonian is essential for a good description of the nuclear structure relevant for the calculation of the NMEs. We use JUN45 effective shell model Hamiltonian \cite{jun45}. Ref. \cite{jun45} provides extensive validation of the JUN45 Hamiltonian by comparing with the experimental data observables such as g.s. and excited states energies, B(E2) values, and magnetic moments. A significant experimental effort was dedicated to containing the nuclear matrix elements by investigating derived observables, such as neutron/proton occupation probabilities \cite{Schiffer2008,Kay2009}, pairing strength, and Gamow-Teller strength \cite{Thies2012}. Here we add to those observables studied in Ref. \cite{jun45} the neutron and proton occupancies in $^{76}$Se and $^{76}$Ge, and the Gamow-Teller strength in $^{76}$Ge.
For the shell model calculations we use the NuShellX@MSU shell-model code~\cite{nushellxmsu}.

Fig. \ref{fig-occn} shows the comparison between our calculated neutron occupancies and the experimental results \cite{Schiffer2008} for the case of $^{76}$Se and $^{76}$Ge. The occupancies of $p_{1/2}$ and $p_{3/2}$ orbital are summed up and denoted with (p). The occupancies of $f_{5/2}$ orbital (f) and of the $g_{p/2}$ orbital (g) are also shown. Fig. \ref{fig-occp} shows the same comparison for the proton occupancies. The data is taken from Ref. \cite{Kay2009}.
We find the agreement between the theoretical results and the experimental data quite satisfactory. 

The validation of the Gamow-Teller strength distribution is particularly relevant for a good description of double beta decay rates. In the $jj44$ valence space the spin-orbit partners orbitals $f_{7/2}$ and $g_{7/2}$ are missing, and the Ikeda sum rule is not satisfied. This results in missing about half of the Gamow-Teller sum-rule, although the loss is at higher energies and is not visible in the low-energy data. A well known problem with the shell model calculation of the Gamow-Teller strength is that the shell model overestimates it, and a quenching factor for the Gamow-Teller operator is necessary to explain the data. For a full major shell valence space, such as as $pf$ model space where all spin-orbit partner orbitals are present, a quenching factor of about 0.74 is validated by the data. In the $jj44$ valence space the violation of the Ikea sum rule requires a modification of this quenching factor. However, the small valence space distorts the high energy strength to lower energy, and for a fine-tuned Hamiltonian such as JUN45, the quenching factor need not be changed too much from its standard value of 0.74. In our case we use a quenching factor of 0.64 that was shown to describe the $2\nu\beta\beta$ NME (see section \ref{decomposition} below).
 
Fig. \ref{fig-gtd} presents the running Gamow-Teller strength for $^{76}$Ge calculated with the JUN45 Hamiltonian and using a quenching factor of 0.64. The horizontal axis represents the excitation energy of the $1^+$ states in the final nucleus $^{76}$As. The results are compared with the high-resolution charge-exchange experimental data \cite{Thies2012}. Although we found discrepancies in the GT strength of individual states of this odd-odd nucleus, $^{76}$As,
the overall theoretical Gamow-Teller strength running sum is in reasonable good agreement with the data.

\section{\onbb NME Results}\label{results}

\subsection{The convergence of the NME}\label{convergence}

First, we studied the convergence properties of the \onbb decay NMEs of $^{76}$Ge. Figure \ref{fig1} presents the total NME (\ref{nme1}) as a function of the number-of-state cutoff parameter $N$ calculated within different  approximations. The red solid line that does not change with $N$ shows the closure NME defined by Eq. (\ref{nme3}). The running closure (\ref{nme5}) and the running nonclosure (\ref{nme4}) NMEs are presented by the red dashed and black dashed curves correspondingly. At large cutoff parameters $N$ the running  NMEs should approach their limits, but it does not occur.
$N\!=\!100$ is the maximum number of states we are able to calculate in $^{76}$As with an computational effort of about 500 000 CPU$\times$h, there is still a significant difference between the running closure and the pure closure values. The mixed matrix elements defined by Eq. (\ref{nme6}) have much better convergence properties, they are presented by the solid black curve on 
Fig. \ref{fig1}. This curve starts with the closure value at $N=0$  
and then slowly increases with $N$ and flattens already after the first 50-60 states. 

\begin{figure}
\begin{center}
\includegraphics[width=0.47\textwidth]{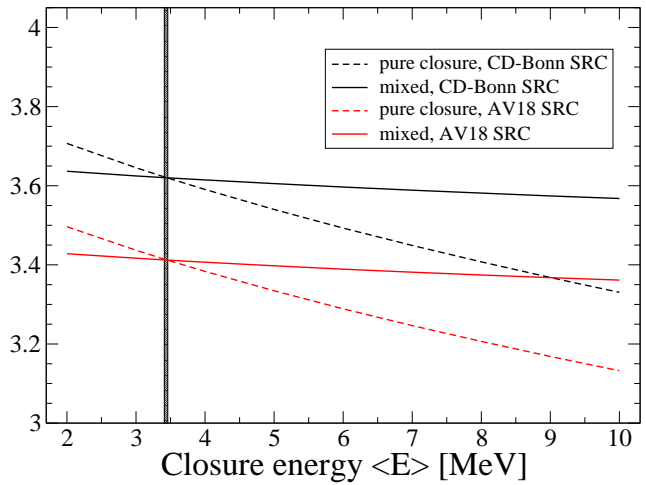}
\caption{ (Color online) Dependence of mixed and closure NMEs for the $0\nu\beta\beta$ decay of ${}^{76}$Ge (light-neutrino exchange) on the average closure energy $\langle E \rangle$. MNEs: closure with CD-Bonn SRC (dashed black curve), mixed with CD-Bonn SRC (solid black curve), closure with AV18 SRC (dashed red curve), and mixed with AV18 SRC (solid red curve).\\}\label{fig3}
\end{center}
\end{figure}


In the mixed method, the states above the cutoff parameter $N$ are included in the closure approximation, which makes the mixed NMEs dependent on the closure energy $\langle E \rangle$. However this dependence is not strong. For $N\!=\!0$ (the closure approximation), it results in a 10\% uncertainty in the total NMEs~\cite{prc10}. When the cutoff parameter increases, this dependence weakens relatively 
rapidly. Figure~\ref{fig2} shows the convergence properties of the mixed NMEs in an enhanced form and how these properties change when the closure energy varies. The solid, dash-dotted, dashed, and dotted lines in the figure present the mixed NMEs calculated with $\langle E\rangle$ equal to 2, 3.4, 7, and 10 MeV, respectively. If we restrict the range of possible closure energies to 3.4 to 7.0 MeV (which is quite reasonable since one curve approaches the final NME from above and the other approaches it from  below, so the true NMEs should be confined somewhere in between), then the corresponding shaded area gives us the uncertainty in the mixed NMEs. We can see how the uncertainty goes down when the cutoff parameter $N$ increases. The corresponding relative error in the mixed matrix elements is presented in the inset in Fig.~\ref{fig2}. It shows that it is sufficient to use only the first 100 nuclear states for each $J^\pi$ of $^{76}$As to obtain the \onbb decay NMEs of $^{76}$Ge within a 1\% accuracy.   

Figures \ref{fig3} shows how the closure NMEs (the dashed curves) and the mixed NMEs calculated with $N=100$ (the solid curves) depend on the closure energy $\langle E \rangle$. There are different ways how the short range correlations (SRC) can be taken into account \cite{prc10}, the upper black curves correspond to the CD-Bonn SRC parametrization set and the lower red curves correspond to the AV18 SRC parametrization set. Fig. \ref{fig3} demonstrates that the mixed NMEs have much weaker dependence on the closure energy than the pure closure NMEs. With the closure energy varying from 2 MeV to 10 MeV the mixed NMEs change by about 2\%, while the closure NMEs change by 12\%. Such observation is consistent with the recent calculations performed for the \onbb decay processes of $^{48}$Ca and $^{82}$Se \cite{sh13,sh14,prc10}.

\subsection{The intermediate $J$ and the $I$-pair decomposition of the NME}\label{decomposition}

\begin{figure}
\begin{center}
\includegraphics[width=0.47\textwidth]{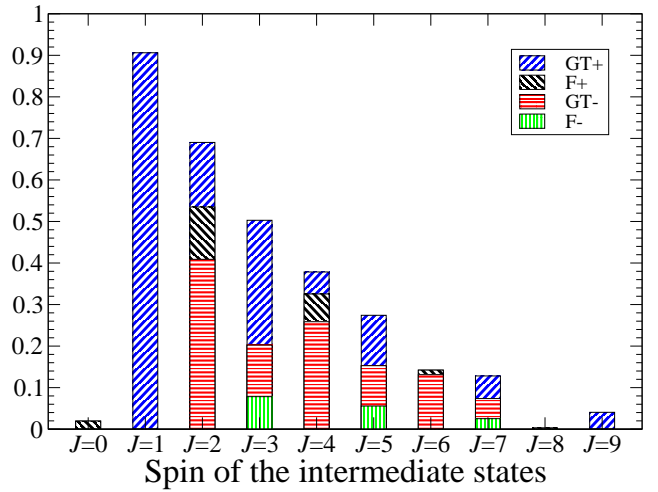}
\caption{(Color online) $J$ decomposition, light-neutrino exchange: contributions of the intermediate states $|\kappa \rangle$ with certain spin and parity $J^\pi$ to the running nonclosure Gamow-Teller (blue and red colors) and Fermi (black and green color) matrix elements for the $0\nu\beta\beta$ decay of ${}^{76}$Ge. Inclined shaded bars correspond to the positive-parity states, while horizontally and vertically shaded bars represent the states with a negative parity. The CD-Bonn SRC parametrization was used.\\}\label{fig4}
\end{center}
\end{figure}

\begin{figure}
\begin{center}
\includegraphics[width=0.47\textwidth]{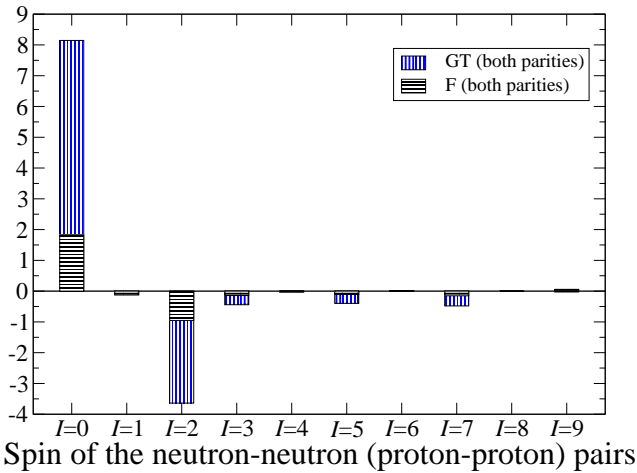}
\caption{(Color online) $I$ decomposition, light-neutrino exchange: contributions to the running nonclosure Gamow-Teller (the blue vertically shaded bars) and Fermi (the black horizontally shaded bars) matrix elements for the $0\nu\beta\beta$ decay of ${}^{76}$Ge from the configurations when two initial neutrons $|24\rangle$ (and two final protons $|13\rangle$) have certain total spin $I$, $\langle 13, I| {\cal O}^\alpha | I, 24\rangle$. The both parities are included. The average energy and the SRC parametrization scheme are the same as in Fig. \ref{fig4}.\\}\label{fig5}
\end{center}
\end{figure}

Figures \ref{fig4} and \ref{fig5} 
present the $J$ decomposition 
[see Eq. (\ref{jdec})] and the $I$ decomposition [see Eq. (\ref{idec})] of the nonclosure NMEs, both figures have similar coloring schemes. For the $J$ decomposition, all Gamow-Teller NMEs with positive (blue inclined shaded bars) and negative (red horizontally shaded bars) parities are positive and all the Fermi matrix elements with positive (black inclined shaded bars) and negative (green inclined shaded bars) parities are negative. Also, all plotted Fermi matrix elements were taken with opposite sign and multiplied by the factor $(g_V/g_A)^2\simeq 0.636$, so if we neglect the Tensor NMEs (which are actually small), then the total height of each bar corresponds to the total NMEs calculated for each spin $J$ in Eq. (\ref{nme1}). We can see that all the spins contribute coherently to the total NMEs. The contribution of $J=1$ is dominating, but it provides only about 30\% of the total value. If we include only the $J=1$ intermediate states, then we will lose about 70\% of the total matrix elements and about 91\% 
of the decay rate. The situation with the $I$ decomposition presented by Figure \ref{fig5} is different. There are big contributions from $I=0$ and $I=2$ which cancel each other. Similar effects have been observed in the shell-model analysis \cite{sh13} for $^{48}$Ca and in \cite{sh14} for $^{82}$Se. Also this $I$ decomposition cancellation was recently discussed in \cite{brown-prl14}, and it was used as a basis for a new method to calculate the NME and to related them to additional nuclear structure constraints that could be obtained form pair transfer reactions \cite{Freeman2007}. 

Table \ref{tbl1} summarizes the results for the light neutrino-exchange NMEs of \onbb decay of $^{76}$Ge calculated within different approximations. The mixed total matrix element is about 7\% percent greater than the total closure NME. This increase is consistent with similar calculations~\cite{sh13,sh14,qrpa-cl}. Table \ref{tbl2} summarizes the results for the light-neutrino-exchange NME \onbb decay of $^{76}$Ge calculated for different SRC parametrization 
sets~\cite{prc10}. 

It should be noted that the $jj44$ model space is incomplete because the $f_{7/2}$ and $g_{7/2}$ orbitals are missing. As a result the Ikeda sum rule is not satisfied and some contributions from the Gamow-Teller NME with $J^\pi=6^+$ and $8^+$ and from the Fermi NME $J^\pi\!=\!1^-$ are missing. Looking at Fig.~\ref{fig4}, it seems safe to suggest that the missing contributions are not very large. However, this deficiency is reflected in the two-neutrino NME, which requires a quenching factor of about 0.64, smaller than the usual 0.74, to describe the experimental data \cite{prl13} (see also Table 2 in Ref.~\cite{caurier12}). Although the spin-isospin operators entering the $0\nu\beta\beta$ decay NME are different from  
those in the pure Gamow-Teller, some authors (see, e.g., Ref.~\cite{ejiri13}) advocate using appropriate quenching factors for contributions coming from different spins of the intermediate states. The most important are those from $J^{\pi}=1^+$ states, which represent about 30\% of the total NMEs, and from $J^{\pi}=2^-$ states \cite{ejiri13}, which represent about 15\% of the total NMEs. It would be interesting to investigate whether quenching factors obtained from other processes, such as $2\nu\beta\beta$ decay and charge-exchange reactions, quench the corresponding contributions to the $0\nu\beta\beta$ decay NMEs. For example, if one uses a quenching factor of $0.64^2$ for the contribution from the $J^{\pi}\!=\!1^+$ states and $0.40^2$ for the contribution from the  $J^{\pi}=2^-$~\cite{ejiri13}, one gets for the CD-Bonn SRC an NME of 2.369 rather than 3.572 (see Table I). One can view this as a lower limit NME in our approach.

\begin{figure}
\begin{center}
\includegraphics[width=0.47\textwidth]{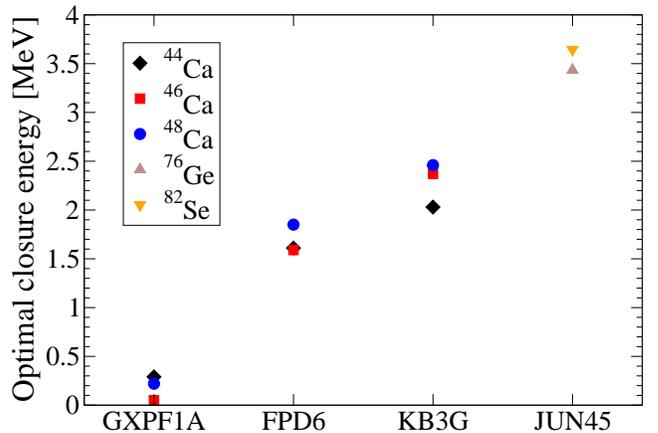}
\caption{ (Color online) Optimal closure energies $\langle E \rangle$ calculated for different isotopes and effective Hamiltonians. Fictitious \onbb decays: $^{44}$Ca (black diamonds) and $^{46}$Ca (red squares). Real decays: $^{48}$Ca (blue circles), $^{76}$Ge (brown upward triangle) and $^{82}$Se (orange downward triangle). Effective Hamiltonians considered are GXPF1A, FPD6, KB3G for Ca and JUN45 for Ge and Se isotopes. \\}\label{fig6}
\end{center}
\end{figure}

\begin{figure}
\begin{center}
\includegraphics[width=0.47\textwidth]{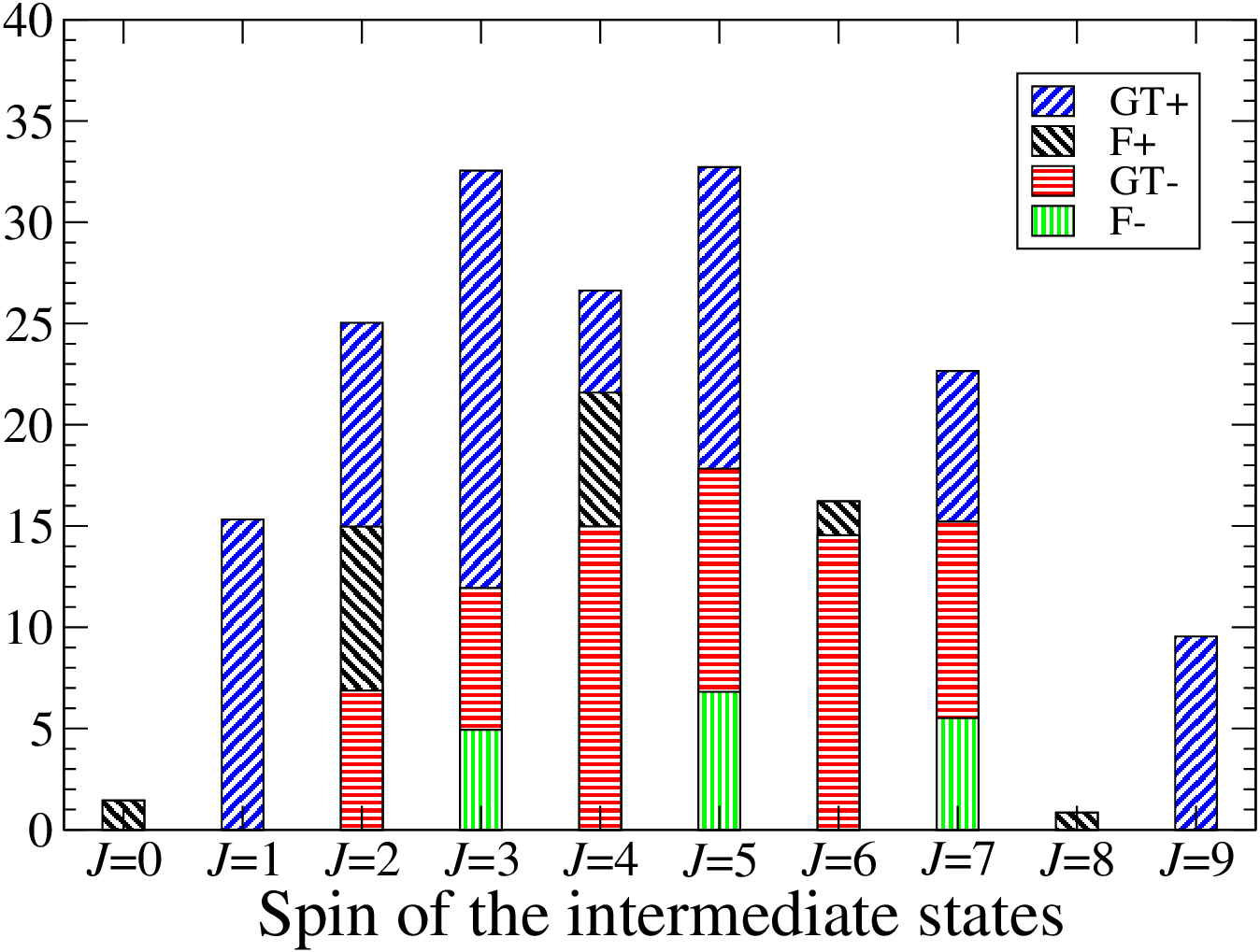}
\caption{ (Color online) $J$ decomposition, heavy-neutrino exchange: contributions of the intermediate states $|\kappa \rangle$ with certain spin and parity $J^\pi$ to the Gamow-Teller (blue and red colors) and Fermi (black and green colors) matrix elements for the \onbb decay of ${}^{76}$Ge. Inclined shaded bars correspond to the contributions of the positive-parity states, while horizontally and vertically shaded bars present the states with a negative parity. All calculations were done with CD-Bonn SRC.\\}\label{fig7}
\end{center}
\end{figure}
 
\begin{table}[ht]
\caption{NMEs for the \onbb decay of $^{76}$Ge  (light-neutrino exchange) calculated within different approximations. All calculations were done 
with CD-Bonn SRC parametrization scheme, the average closure energy $\langle E \rangle=9.41$ MeV \cite{tomoda}. \\}\label{tbl1}
\begin{ruledtabular}
\begin{tabular}{lcccc}
 & Closure & Run.Closure & Run.Nonclosure & Mixed \\
\hline
$M^{0\nu}_{GT}$ & 2.95 & 2.50 & 2.70 & 3.15 \\
$M^{0\nu}_F$ & $-0.65$ & $-0.58$ & $-0.61$ & $-0.67$ \\
$M^{0\nu}_T$ & $-0.01$ & 0.02 & 0.02 & $-0.01$ \\
$M^{0\nu}_{total}$ & 3.35 & 2.89 & 3.10 & 3.57 \\ 
\end{tabular}
\end{ruledtabular}
\end{table}

\begin{table}[ht]
\caption{Mixed and pure closure (last column) NMEs for the \onbb decay of $^{76}$Ge (light-neutrino exchange) calculated with different SRC parametrizations schemes \cite{prc10}. Closure NMEs were calculated for a standard average closure energy of $\langle E \rangle=9.41$ MeV \cite{tomoda}.\\}\label{tbl2}
\begin{ruledtabular}
\begin{tabular}{lccccc}
SRC & $M^{0\nu}_{GT}$ & $M^{0\nu}_F$ 
& $M^{0\nu}_T$ & $M^{0\nu}_{total}$ & ${\cal M}^{0\nu}_{closure}$\\
\hline
None & 3.06 & $-0.63$ & $-0.01$ & 3.45 & 3.24 \\
Miller-Spencer &  2.45 & $-0.44$ & $-0.01$ & 2.72 & 2.55 \\
CD-Bonn & 3.15 & $-0.67$ & $-0.01$ & 3.57 & 3.35 \\
AV18 & 2.98 & $-0.62$ & $-0.01$ & 3.37 & 3.15 \\
\end{tabular}
\end{ruledtabular}
\end{table}

\begin{figure*}
\centering
\begin{minipage}[t]{.44\textwidth}
\includegraphics[width=1.00\linewidth]{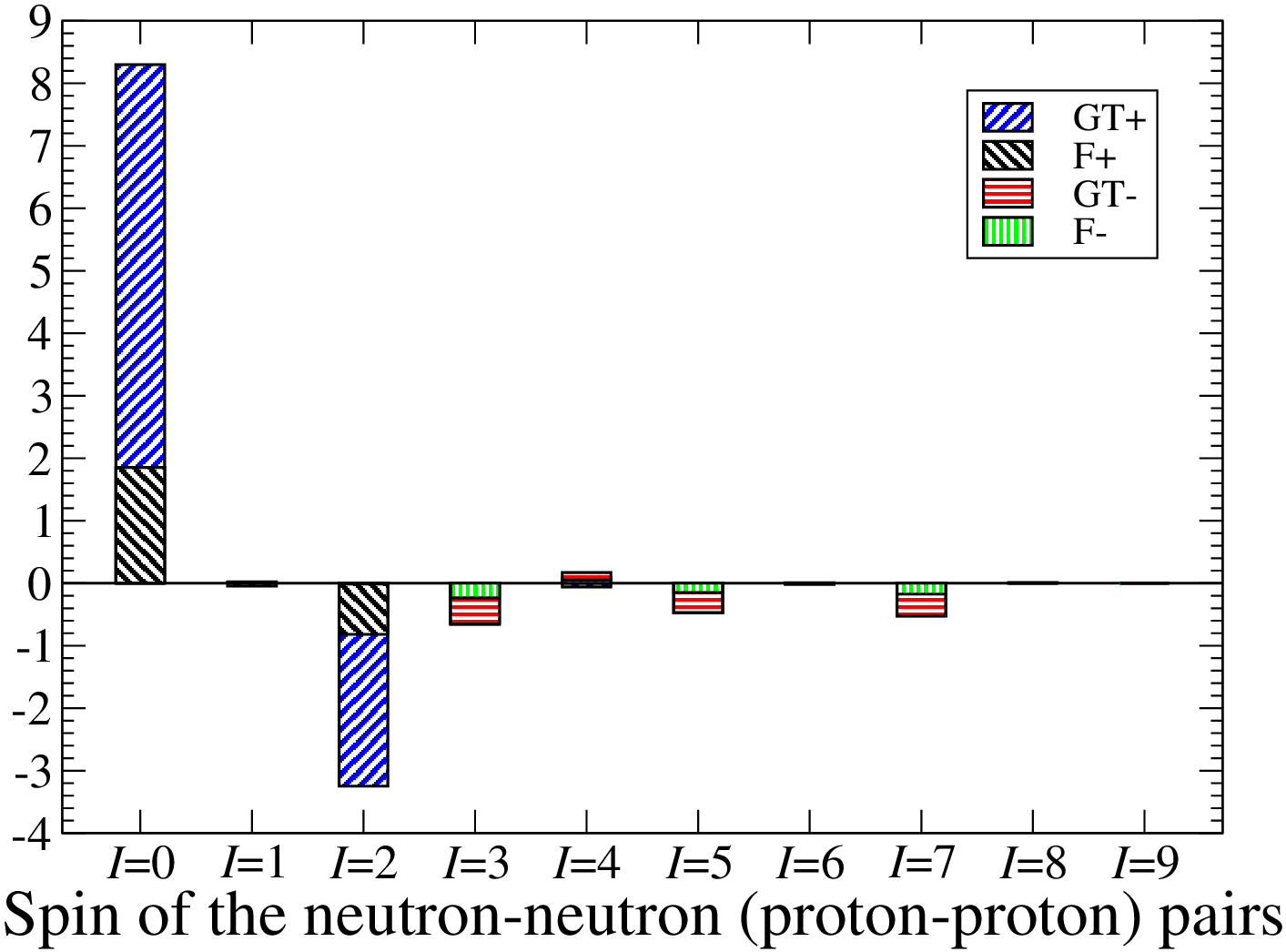}
\caption{(Color online) $I$ decomposition: closure approximation Gamow-Teller and Fermi matrix elements (both parities) for the \onbb decay of $^{76}$Ge, light-neutrino exchange. The calculation performed with the optimal closure energy, $\langle E \rangle=3.5$ MeV. The results should be compared with the matrix elements presented on Fig. \ref{fig5}.}\label{fig8}
\end{minipage}\qquad
\begin{minipage}[t]{.44\textwidth}
\includegraphics[width=1.00\linewidth]{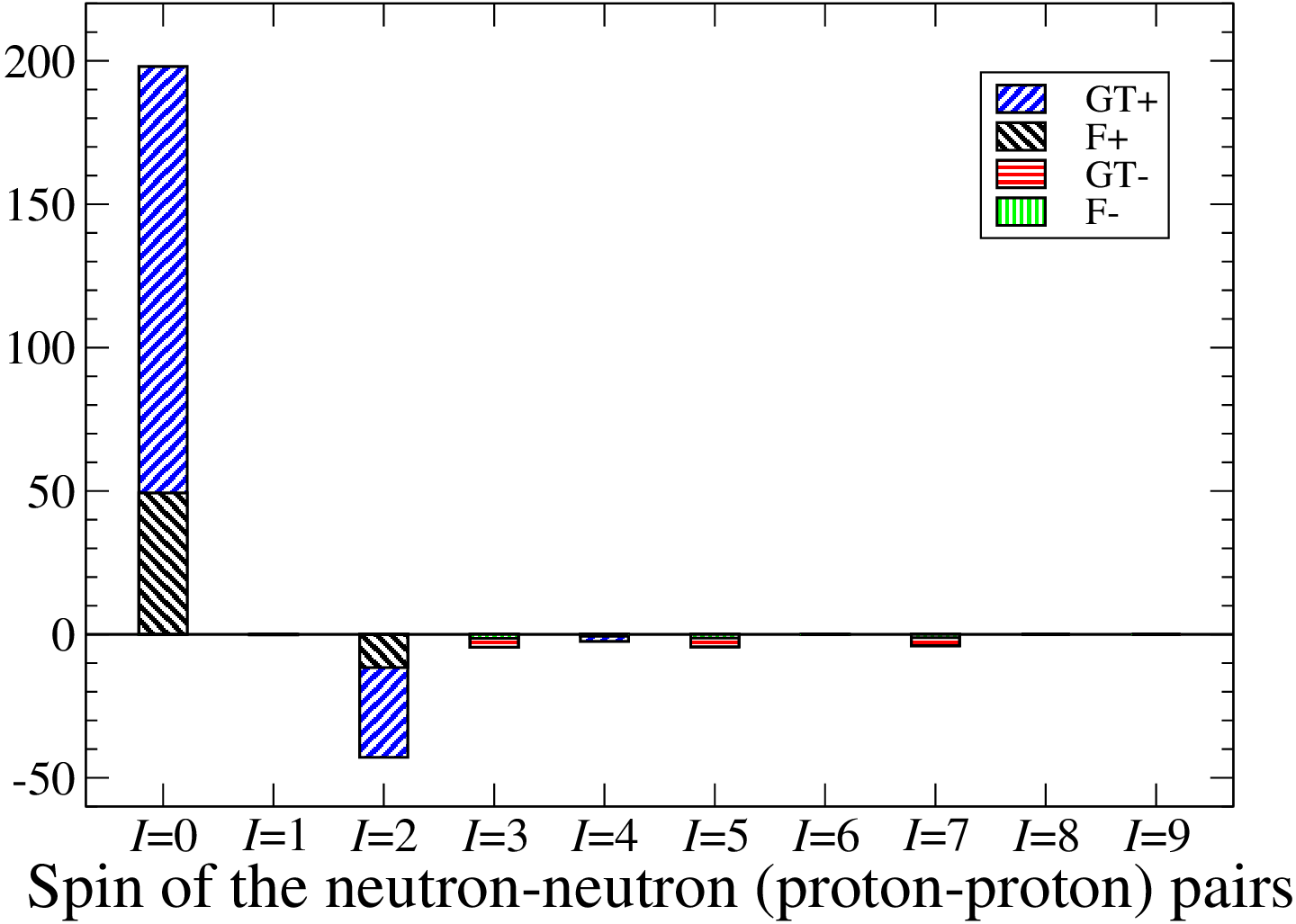}
\caption{(Color online) $I$ decomposition: closure approximation Gamow-Teller and Fermi matrix elements (both parities) for the \onbb decay of $^{76}$Ge, heavy-neutrino exchange. }\label{fig9}
\end{minipage}
\end{figure*}

\begin{table*}[ht]
{\caption{Comparison of the total NMEs for the \onbb decay of $^{76}$Ge (light-neutrino exchange) calculated with different approaches and with different SRC parametrizations schemes. 
$g_A=1.254$ is used for the axial-vector 
coupling constant.\\}\label{tbl3}}
\begin{ruledtabular}
\begin{tabular}{llcccccccc}
\rule{0cm}{0.33cm} 
& & ISM & ISM & QRPA(TBC) & RQRPA(TBC) & QRPA(J) & QRPA & IBM-2 & EDF \\
& SRC & present & \cite{npa818} & \cite{qrpa-tbc1,qrpa-tbc2} & \cite{qrpa-tbc1,qrpa-tbc2} & \cite{suh12} & \cite{engel13} & \cite{iba-2} & \cite{gcm} \\
\hline
$M_{total}^{0\nu}$, & \mbox{None} & 
3.45 & 2.96 &  &  &  &  & & \\
 & \mbox{Miller-Spencer} & 
2.72 & 2.30 & 4.68 & 3.33 & 3.77 & 3.83 & 5.42 & \\
 & \mbox{CD-Bonn} & 
3.57 &      & 6.32 & 5.44 &  &    & 6.16 &  \\
 & \mbox{AV18} & 
3.37 &      & 5.81 & 4.97 &   &   & 5.98 &  \\
 & \mbox{UCOM} & 
     & 2.81 & 5.73 & 3.92 & 5.18 &   &   & 4.60 \\
\end{tabular}
\end{ruledtabular}
\vspace*{0.5cm}
\end{table*}

\subsection{The optimal closure energy}\label{optimal}

Since we can calculate both the nonclosure NME and the closure NME, 
it is possible to find such optimal values for the closure energies at which the closure approach provides the most accurate NMEs (see, e.g., the crossing lines in Fig. \ref{fig3}):
\beq \label{eq15}
{\bar M}^{0\nu}={\cal M}^{0\nu}(\langle E\rangle).
\eeq
One interesting observation is that the optimal energies calculated for the \onbb decay of $^{82}$Se \cite{sh14} and $^{76}$Ge with the same JUN45 effective Hamiltonian and the same $jj44$ model space 
practically coincide: they both equal about $\langle E \rangle \approx 3.5$ MeV, although the two cases describe quite different nuclei. It would thus be interesting to find a method to estimate the optimal closure energies rather then using estimates from other methods, such as those in Ref.~\cite{tomoda}. Figure~\ref{fig6} presents the optimal closure energies calculated for the fictitious \onbb decays of $^{44}$Ca (diamonds) and $^{46}$Ca (squares) and for the realistic \onbb decays of $^{48}$Ca (circles), $^{76}$Ge (upward triangles), and $^{82}$Se (downward triangles). All calcium 
isotopes were calculated in the $pf$ model space using several 
realistic Hamiltonians. The $^{76}$Ge and $^{82}$Se isotopes were 
considered in the same $jj44$ model space and with the same JUN45 
Hamiltonian. The optimal closure energies are significantly lower than the standard closure energies (7.72 MeV for Ca, 9.41 MeV for Ge, and 10.08 MeV for Se~\cite{tomoda}), which explains the 7--10\% growth in absolute values of the nonclosure NMEs compared to the closure values. We conjecture that the optimal energies depend on the effective Hamiltonian and, possibly, on the model space. We found the optimal closure energies for the three Hamiltonians in the $pf$ model space: GXPF1A~\cite{gxpf1a}, FPD6~\cite{fpd6}, and KB3G~\cite{kb3g}. 
However, it seems that the energies do not depend much on the specific nucleus: all the calcium isotopes calculated with the same Hamiltonian and both the $^{76}$Ge and the $^{82}$Se isotopes calculated with the same model space and with the same Hamiltonian give similar optimal closure energies. This opens up an interesting opportunity: one could calculate the optimal closure energy in a realistic model space with an effective Hamiltonian for a nearby less computationally demanding isotope (for example, $^{44}$Ca), after which one could use it for a realistic case (for example, $^{48}$Ca). 
This scheme offers a consistent way of ``calculating" the closure 
energies that has not been discussed before. In the Table \ref{tbl3} we compare our results for the NMEs of \onbb decay of $^{76}$Ge (light-neutrino exchange mechanism) with the recent calculations. Table \ref{tbl3} presents matrix elements obtained with: interacting shell model approach (ISM) \cite{npa818}; quasiparticle random phase approximation, T{\"u}ebingen-Bratislava-Caltech group [(R)QRPA(TBC)] \cite{qrpa-tbc1,qrpa-tbc2}; quasiparticle random phase approximation, Jyv{\"a}skyl{\"a} group [QRPA(J)] \cite{suh12}; quasiparticle random phase approximation, Holt and Engel \cite{engel13}; interacting boson model (IBM-2) \cite{iba-2}; and generator coordinate method (EDF) \cite{gcm}. The value $g_A=1.254$ is used in most of the calculations, except for IBM-2, which uses the axial-vector coupling constant $g_A=1.269$ \cite{iba-prl}. 

\subsection{The heavy neutrino-exchange NME}\label{heavy}

Figure \ref{fig7} and Table \ref{tbl4} summarize the results for our 
heavy-neutrino exchange \onbb decay of $^{76}$Ge. Comparing Figs. \ref{fig4} and \ref{fig7} we can see that the heavy neutrino-exchange NMEs do not vanish with the large intermediate spins $J$. The heavy-neutrino potentials have a strong short-range part, so the contributions from the large neutrino momentum, which are responsible for the higher spin contributions, are not suppressed.    

\begin{table}[ht]
\caption{Heavy neutrino-exchange NMEs of the \onbb decay of $^{76}$Ge 
calculated with different SRC parametrizations sets \cite{prc10}.   \\}\label{tbl4}
\begin{ruledtabular}
\begin{tabular}{lcccc}
SRC, Approximation  & $M^{0\nu}_{GT}$ & $M^{0\nu}_F$ & 
$M^{0\nu}_T$ & $M^{0\nu}_{total}$ \\
\hline
CD-Bonn, Closure & 162 & $-62.6$ & $-0.19$ & 202 \\
CD-Bonn, Run.Closure & 147 & $-56.5$ & 0.22 & 183 \\
AV18, Closure & 105 & $-52.1$ & $-0.20$ & 140 \\
AV18, Run.Closure & 95.8 & $-46.9$ & 0.22 & 126 \\
\end{tabular}
\end{ruledtabular}
\end{table}

\subsection{The $I$ decomposition of the closure NME}\label{Idecomposition2}

Finally we calculated $I$ decompositions of the closure NMEs, Eq. (\ref{nme3}), for the \onbb decay of $^{76}$Ge at the optimal closure energy calculated specifically for $^{76}$Ge, for the JUN45 effective Hamiltonian and the jj44 model space, $\langle E \rangle = 3.5$ MeV. Figs. \ref{fig8} and \ref{fig9} present the matrix elements calculated for the light-neutrino and heavy-neutrino exchanges correspondingly. NMEs on these figures include both, positive and negative, and the Fermi matrix elements were taken with the opposite sign and multiplied by a factor of $(g_V/g_A)^2$, so that the total hight of each bar corresponds to the total matrix element (\ref{nme1}) (if the tensor matrix element is neglected). Comparing Fig. \ref{fig5} and Fig. \ref{fig8} we can see a good agreement between the nonclosure and the closure approximations when the optimal closure energy is used. It is important to note that using optimal closure energy for the closure NMEs provides good results not only for the total matrix element but also for the individual MNEs, of different types and different spins $I$.   

\section{Conclusions and Outlook}

In summary, we calculated the \onbb decay NME of $^{76}$Ge using, for the first, time a realistic shell-model approach beyond closure approximation. For the calculation we used the realistic $jj44$ model space and the JUN45 effective Hamiltonian that was fine tuned in the region of $^{76}$Ge and $^{82}$Se. We investigated a new method, which considers information from both closure and nonclosure approaches. This mixed method was carefully tested on the fictitious cases of $^{44}$Ca and $^{46}$Ca where all the intermediate sates can be calculated. Then the mixed method was used to calculate the \onbb decay NMEs of $^{48}$Ca, $^{82}$Se, and $^{76}$Ge isotopes, which was the first realistic shell-model calculation of the \onbb decay NMEs beyond closure approximation. We demonstrated that the NMEs calculated with the mixed method converge very rapidly compared to the running nonclosure matrix elements and we found a 7-10\% increase in the total NMEs compared to the closure values. 

For the light-neutrino-exchange mechanism we predict 
\beq \label{res}
M^{0\nu} = 3.5 \pm 0.1,
\eeq
where the average value and the error were estimated considering the total mixed NMEs from Table \ref{tbl2} calculated with CD-Bonn and AV18 SRC parametrization sets.  A more elaborate method of estimating the error, which rely in part on our $I$-pair decomposition, is presented in Ref. \cite{BrownFangHoroi2015}. For the heavy-neutrino exchange NME we get with different SRC parametrization sets (CD-Bonn and AV18 SRC): 
\beq \label{res2}
M^{0\nu}_N = 202/140.
\eeq

We proposed a new method of calculating the optimal closure energies with which the closure approach gives the most accurate NMEs. We argue that these optimal closure energies depend on the Hamiltonian and model space and have a weak dependence on the actual isotopes. This features can be used to determine the optimal closure energies using fictitious double-$\beta$ decay of isotopes that
are easier to calculate in a given valence space. This computational route offers the opportunity of estimating the beyond-closure \onbb NMEs without actually calculating the intermediate states.

We calculated for the first time a decomposition of the shell-model NMEs in light and heavy neutrino-exchange mechanisms for different spins of intermediate states. We found that for the light-neutrino-exchange NMEs the contribution of the $J^{\pi}=1^+$ states is about 30\% and that of the $J^{\pi}=2^-$ states is about 15\%.  
The shell-model $J$ decomposition that we obtained provides a unique 
opportunity to selectively quench different contributions to the total NMEs, which, in the case of $^{76}$Ge, could lead to a decrease in the total matrix elements by about 30\%. Although the QRPA approach can provide a $J$ decomposition, its methodology of choosing the $g_{pp}$ parameter to describe the $2\nu\beta\beta$ half-life~\cite{qrpa-cl} could make the selective quenching ambiguous.

We also presented $I$-pair decompositions for both light and heavy neutrino exchange NMEs that were recently used as a starting point to
propose a new method of calculating these matrix elements \cite{brown-prl14}, and which could lead to new venues of constraining the NME by pair transfer experimental data. In addition, the different levels of cancellation between $I=0$ and $I=2$ contributions could shed new light on the origin of the discrepancies between NME calculated with different methods  \cite{BrownFangHoroi2015}.

\bigskip

The authors thank B.A. Brown and V. Zelevinsky for useful discussions. Support from  the NUCLEI SciDAC Collaboration under
U.S. Department of Energy Grant No. DE-SC0008529 is acknowledged. 
M.H. also acknowledges U.S. NSF Grant Nos. PHY-1068217 and PHY-1404442.

\end{document}